\newcommand\pa{\partial}

\newcommand{\beque}{\begin{equation*}}
\newcommand{\eeq}{\end{equation}}
\newcommand{\beq}{\begin{equation}}
\newcommand{\eeque}{\end{equation*}}
\newcommand{\beqnl}{\begin{eqnarray}}
\newcommand{\eeqna}{\end{eqnarray*}}
\newcommand{\beqna}{\begin{eqnarray*}}
\newcommand{\eeqnl}{\end{eqnarray}}

\newcommand\lagr{{\mathcal L}}

\newcommand{\sgn}{\operatorname{sgn}}

\providecommand{\sign}[1]{\textrm{sign{$#1$}}}

\documentclass[aps,prd]{revtex4}

\usepackage{amsfonts}
\usepackage{amsthm}
\usepackage{amsbsy}
\usepackage{amssymb}
\usepackage{amsmath}
\usepackage{graphicx}
\usepackage{color}

\begin{document}

\date{\today}

\title{The Hawking effect in dielectric media and the Hopfield model}

\author{F.~Belgiorno$^1$, S.L.~Cacciatori$^{2,3}$, F.~Dalla~Piazza$^4$}

\address{$^1$ Dipartimento di Matematica, Politecnico di Milano, Piazza Leonardo 32, IT-20133
Milano, Italy, and INdAM, GNFM, Italy\\
$^2$ Department of Science and High Technology, Universit\`a dell'Insubria, Via Valleggio 11, IT-22100 Como, Italy\\
$^3$ INFN sezione di Milano, via Celoria 16, IT-20133 Milano, Italy\\
$^4$ Universit\`a ``La Sapienza'', Dipartimento di Matematica, Piazzale A. Moro 2, I-00185, Roma,   Italy}

\begin{abstract}

We consider the so-called Hopfield model for the electromagnetic field in a dielectric
dispersive medium in a framework in which one allows a space-time dependence of microscopic parameters,  
aimed to a phenomenological description of a space-time varying dielectric perturbation
induced by means of the Kerr effect. We discuss the analogue Hawking effect 
{\color{black} by introducing a simplified model which 
avoids some difficulties which characterize in the full Hopfield model, still keeping 
the same dispersion relation. Our main result is an analytical calculation of  
the spontaneous thermal emission in the single-branch case, which is provided non perturbatively for the first time in the framework of dielectric black holes. An universal mechanism for thermality which is shared both by optical black holes and acoustic black holes is also pointed out.} 

\end{abstract}

\maketitle

\section{Introduction}

In the framework of analogue gravity, a very interesting possibility to check experimentally the 
existence of the analogue of Hawking radiation is represented by black holes in dielectric media. 
By taking into account in a very short summary the historical developments, we point out that 
there are several contributions to this topic, which was reassessed in a interesting framework in 
\cite{philbin-leonhardt} and received a further impulse, especially on the experimental side, in 
\cite{faccio-belgiorno-prl}, grounded on \cite{belgiorno-prd,cacciatori-njp}, and then corroborated also 
by numerical simulations in \cite{rubino-njp}. Experimental results in \cite{faccio-belgiorno-prl} are controversial 
(see e.g. \cite{unschu,faccio-prl-ans,unschu-prd,prain}). Subsequent studies involve both numerical and 
experimental situations. On the former side, an interesting 2D model where there is a sharp step behavior of the dielectric properties 
of the medium has been discussed in \cite{finazzi-carusotto-kin,
finazzi-carusotto-pra13,finazzi-carusotto-pra14}, with numerical evaluations of the pair-production 
processes involved. Numerical but also analytical perturbative studies appeared in \cite{petev-prl,rubino-sr}, 
where a smooth behavior of the refractive index was assumed, in the framework of a phenomenological model 
grounded on the UPPE approximation for the electromagnetic field, 
and thermality has been explored again in particular in \cite{petev-prl}. The aforementioned studies 
contributed in a very important way to the comprehension of the physics at hand. 
In the present paper, 
our {\color{black} main contribution is to complement these studies on the analytical side, providing both a theoretical framework 
and analytical nonperturbative deduction of thermality which has been lacking up to now for dielectric black holes. Our reference model is the covariant Hopfield model discussed in \cite{belcacciadalla-pert,belcacciadalla-hopfield}. It represents our basic tool  for  analyzing conceptual issues 
which characterize the physics at hand. We then introduce a simplified model for dispersion, which is still 
covariant, but has the great advantage to be not involved in a constrained quantization procedure and also 
to be based on a couple of scalar fields (in place of a couple of vector fields). 
Its physical content is nontrivial, and can allow for a number of very interesting physical situations, 
which can be also experimentally tested. We stress that the full Hopfield model can also be exactly solved in 
the same physical situations, with the difference that it is much more difficult to be handled, both for the 
vectorial nature of the fields involved, and for their being constrained systems.\\
The main focus of our paper is in the analytical deduction of thermality for dielectric black holes. 
This represents our key-result, and main motivation of our analysis. 
Furthermore, we point out a peculiar Fuchsian singularity structure of the field equations 
near the horizon, and stress its role in determining thermality of the emitted radiation, and is shared with 
other analytical calculations in the acoustic black hole case \cite{corley,cpf}.  As a by-product of our analysis, we provide also a derivation of the so-called generalized Manley-Rowe relations, which 
play an important role in presence of pair-creation. See Appendix \ref{manley}.}\\

{\color{black} The plan of the paper is the following.} In sec. \ref{hopf-model} we start from 
a short discussion of the Hopfield model, {\color{black} and describe a simplified model which 
avoids some formal difficulties of the Hopfield model, still maintaining the same dispersion 
relation.}
 We 
develop the scattering picture in order to point out the presence of pair creation from the vacuum. 
Then in sec. \ref{thermality} we 
derive thermality in the limit of weak dispersion by matching asymptotic solutions with solutions 
obtained in the near horizon region, in strict analogy with calculations developed in \cite{corley,cpf}.  
In sec. \ref{horizons} we discuss the problem of the nature of the horizon involved in the pair creation 
process. In sec. \ref{multibranches} we adapt our results to the case of multiple resonances in the 
dielectric medium. Then in sec. \ref{conclusions} we summarize our main results. For the sake of completeness, 
we also provide some appendices.  In appendix 
\ref{exact} we show that our model  reproduces  
exactly the same dispersion relation as in the full electromagnetic case. Moreover, we show that, in line of 
principle, the model can be solved exactly, being reducible to a Gaussian path integral. 
{\color{black} In appendix \ref{scattering-asympt} we discuss some formal aspects of scattering theory. 
In appendix \ref{manley} we derive the generalized Manley-Rowe relations in our scattering framework. 
In appendix \ref{geomopt}, some discussion of geometrical optics aspects is given.}

\section{Hopfield model revisited: the $\varphi \psi$--model as  a theoretical benchmark for the electromagnetic case}
\label{hopf-model}

The Hopfield model is aimed to the description of dielectric dispersive media, considered as transparent 
\cite{Hopfield,Fano}. 
{\color{black} The model, as known, introduces mesoscopic fields, representing the polarization fields of mesoscopic electrodynamics, 
which are coupled with the electromagnetic 
field. These fields are 
in form of a number of harmonic oscillators, each with a proper oscillation frequency characterizing 
dispersive effects of light in the medium. The medium itself is considered in first approximation 
as transparent.}
The latter requirement can be relaxed, as e.g. in \cite{suttorp,suttorp-jpa,suttorp-wubs}. 
In particular, in \cite{suttorp-jpa} an interesting model including space-time variations of the susceptibility is 
provided. The original model can be described as 
follows:
\beqna
\lagr_{em}:&=&\frac{1}{8\pi} \left( \frac{1}{c} \dot{\mathbf{A}} +\nabla \phi\right)^2
-\frac{1}{8\pi} \left(\nabla\wedge \mathbf{A}\right)^2\cr
&+&\frac{1}{2\chi} \left(\dot{\mathbf{P}}^2 -\omega_0^2 \mathbf{P}^2\right)
-\frac{1}{2 c} \left(\mathbf{P}\cdot \dot{\mathbf{A}}+\dot{\mathbf{A}}\cdot \mathbf{P}\right)
-\frac{1}{2} \left(\mathbf{P}\cdot \nabla \phi+\nabla \phi\cdot \mathbf{P}\right).
\eeqna
{\color{black} In the previous equation, 
$\mathbf{A},\phi$ stay for the components of the electromagnetic potential vector field, $\mathbf{P}$ is the 
polarization field, $\omega_0$ is the proper frequency for the 
oscillation of the polarization field, $\chi$ is the dielectric susceptibility, and $c$ is the speed of light}. 
We introduced in \cite{belcacciadalla-hopfield,belcacciadalla-pert} a full covariant 4D version of the Hopfield model, 
taking into account the requirements for a relativistic extension, which is apt for a 
simple discussion of physics in the comoving frame of a uniformly moving 
dielectric perturbation induced by means of the Kerr effect. 
{\color{black} This is only a phenomenological approximation, in the sense that, in place of  
the nonlinear interactions involved in the Kerr effect, we consider its effects through 
the space-time dependence of parameters like the susceptibility. See below.}\\
The covariant lagrangian density deduced in \cite{belcacciadalla-hopfield} is 
\beqnl
\lagr = &&-\frac{1}{16\pi} F_{\mu \nu} F^{\mu \nu}
-\frac{1}{2\chi \omega_0^2} \left[ (v^\rho \pa_\rho P_\mu) (v^\sigma \pa_\sigma P^\mu) \right]\cr
&&
+ \frac{1}{2\chi}  P_\mu P^\mu
-\frac{g}{2 c} (v_\mu P_\nu- v_\nu P_\mu) F^{\mu \nu},
\eeqnl
{\color{black}  where $P_\mu$ is the polarization field, $F_{\mu \nu}$ is the field strength tensor, and $v^\mu$ is the four velocity associated with the dielectric medium.} 
In this model, we allow the microscopic parameters $\chi,\omega_0,g$ to depend on 
space-time variables. We notice that, as e.g. in the (non covariant) polariton model introduced in \cite{barnett,suttorp}, 
where a space dependent coupling between the electromagnetic field and the polarization field is 
introduced, we have as well introduced a coupling $g$, which is a priori space-time dependent.\\
The covariant quantization of the model has been pursued in \cite{belcacciadalla-hopfield}, 
and is a nontrivial task because of the gauge constraints implied by the model. See also 
\cite{belcacciadalla-pert}, where further analysis is presented in a perturbative framework. In order 
to look for our main goal, which is the Hawking effect in dielectric media, we simplify as 
possible the theoretical framework, as discussed in the following.\\ 


The electromagnetic lagrangian for the full Hopfield model is quite involved, and in order to 
test quantum effects is not so manageable. In order to gain insights into the real situation, 
and carry out analytical calculations as far as possible, we introduce a simplified model where 
the couple {\sl electromagnetic field - polarization field} is simulated by a {couple of scalar fields: 
$\varphi$, $\psi$} in place of the electromagnetic field and of the polarization field respectively. The 
model we propose is constructed in such a way to maintain the same dispersion relation and to simulate 
the same coupling as in the full case. The model at hand is related to the 2D reduction of the Hopfield model 
which is adopted in \cite{finazzi-carusotto-pra13}.\\  
We introduce
\beq
{\cal L}_{\varphi\psi} = \frac{1}{2} (\partial_\mu \varphi)(\partial^\mu \varphi)+
\frac{1}{2\chi \omega_0^2} \left[ (v^\alpha \partial_\alpha \psi)^2 - \omega_0^2 \psi^2 \right] + \frac{g}{c} (v^\alpha \partial_\alpha \psi) \varphi,
\label{lagrangian}
\eeq
where $\chi$ plays the role of the dielectrical susceptibility, $v^\mu$ is the four-velocity vector, $\omega_0$ stays for the proper frequency  of the medium, and $g$ is 
a constant which plays the role of coupling constant between the fields. A priori, in a phenomenological 
model aimed at describing the electromagnetic field, we can leave room for a space-time dependence of microscopic parameters 
like $\chi,\omega_0,g$. Moreover, we can extend the model in such a way to include also $N>1$ polarization fields $\psi_i$, 
each one characterized by a different $\omega_{0i},\chi_i,g_i$. We shall not use the latter freedom till the final sections 
of the paper, and focus our attention on the single-resonance model with just a single polarization field.\\

The equations of motion are (we omit space-time arguments):
\beqnl
&& \square \varphi - \frac{g}{c}  (v^\alpha \partial_\alpha \psi)=0,\\
&& (v^\alpha \partial_\alpha) \frac{1}{\chi \omega_0^2} (v^\alpha \partial_\alpha \psi) + \frac{1}{\chi} \psi + 
\frac{1}{c} v^\alpha \partial_\alpha  ( g \varphi)=0. 
\eeqnl
Let $G_\psi$ be the Green function for $\psi$: 
\beq
[ (v^\alpha \partial_\alpha) \frac{1}{\chi \omega_0^2} (v^\alpha \partial_\alpha) + \frac{1}{\chi}]G_\psi = \delta,
\eeq
where $\delta$ stays for the Dirac delta function. Then, defining $\tilde{x}=(t,x,y,z)$, 
we get the following system: 
\beqnl
&& \psi (\tilde{x}) =  \frac{1}{c} \int d^4w G_\psi (\tilde{x}-w)  (v^\alpha \partial_\alpha g\varphi) (w), \label{eq-psi}\\
&& \square \varphi (\tilde{x}) - \frac{g}{c^2}  (v^\alpha \partial_\alpha ) \int d^4w G_\psi (\tilde{x}-w)  (v^\beta \partial_\beta g \varphi) (w)=0, 
\label{eq-varphi}
\eeqnl
which represent a simplified model equations set simulating dispersive effects in optics.\\
It is also useful to introduce the Hamiltonian equations for the given model. In particular, we 
introduce the conjugate momenta
\beqnl
\pi_\varphi  &:=& \frac{\partial {\cal L}}{\partial \partial_0 \varphi}=\partial_0 \varphi, \label{mo-phi}\\
\pi_\psi &:=& \frac{\partial {\cal L}}{\partial \partial_0 \psi}=\frac{1}{\chi \omega_0^2} v^0 v^\alpha 
\partial_\alpha \psi + \frac{g}{c} v^0 \varphi \label{mo-psi}.
\eeqnl
Then we calculate the Hamiltonian density ${\cal H}$:
\beqnl
{\cal H} &=& \partial_0 \psi \pi_\psi + \partial_0 \varphi \pi_\varphi - {\cal L} \cr
&=& \frac 12 \pi_\varphi^2 + \frac{\chi \omega_0^2}{2 (v^0)^2} \pi_\psi^2 -\frac{v^k}{v^0} (\partial_k \psi) \pi_\psi 
-\frac{g}{c} \frac{\chi\omega_0^2}{v^0}  \pi_\psi \varphi + \frac{1}{2c^2} \chi \omega_0^2 g^2 \varphi^2 +\frac{1}{2\chi} \psi^2 +\frac 12 (\partial_k  
\varphi)^2.
\eeqnl
Let us also define $h :=\int dx {\cal H}$. Then we obtain 
\beqnl
\partial_0 \varphi &=& \{\varphi,h\},\\
\partial_0 \psi &=& \{\psi,h\},\\
\partial_0 \pi_\varphi &=& \{\pi_\varphi,h\},\\
\partial_0 \pi_\psi &=& \{\pi_\psi,h\}.
\eeqnl
Of course, the following non trivial Poisson brackets hold true:
\beqnl
\{\varphi,\pi_\varphi\} &=& \delta ,\\
\{\psi,\pi_\psi\} &=& \delta,
\eeqnl
which will play a role in the quantization of the fields, as in the Hopfield model \cite{belcacciadalla-hopfield}.

\subsection{Conserved scalar product}

By following the same line of thought as in the electromagnetic case, we can show that the following (global) phase 
transformation
\beqnl
&& \varphi \mapsto e^{i a} \varphi; \quad \quad \varphi^* \mapsto e^{-i a} \varphi^*,\\
&& \psi \mapsto e^{i a} \psi; \quad \quad \psi^* \mapsto e^{-i a} \psi^*,
\eeqnl
is a symmetry for the complexified Lagrangian
\beq
{\cal L}^{complex}_{\varphi\psi} = \frac{1}{2} (\partial_\mu \varphi^*)(\partial^\mu \varphi)+
\frac{1}{2\chi \omega_0^2} \left[ (v^\alpha \partial_\alpha \psi^*)(v^\alpha \partial_\alpha \psi)
 + \omega_0^2 \psi^2 \right] + \frac{g}{2c} (v^\alpha \partial_\alpha \psi^*) \varphi + 
 \frac{g}{2c}  (v^\alpha \partial_\alpha \psi) \varphi^*;
\eeq
Noether's theorem implies that the following current:
\beq
J^\mu:= \frac{i}{2} \left[ \varphi^* \partial^\mu \varphi - (\partial^\mu \varphi^*) \varphi 
+ \frac{1}{\chi \omega_0^2} v^\mu \psi^* v^\alpha \partial_\alpha \psi - 
\frac{1}{\chi \omega_0^2} v^\mu \psi v^\alpha \partial_\alpha \psi^* +\frac{g}{c}  v^\mu (\psi^* \varphi - \psi \varphi^*)\right] 
\label{curr-j}
\eeq
is conserved: $\partial_\mu J^\mu=0$ along the solutions of the equations of motion. Then the charge 
\beq
\int_{\Sigma_t} dx J^0
\eeq
is conserved, and allows us to define a conserved scalar product. In particular, we obtain 
\beq 
((\varphi\ \psi)|(\tilde{\varphi}\ \tilde{\psi})) =  \frac{i}{2} \int_{\Sigma_t} \left[ \varphi^* \partial^0 \tilde{ \varphi} - 
(\partial^0 \varphi^*) \tilde{\varphi} 
+ \frac{1}{\chi \omega_0^2} v^0 \psi^* v^\alpha \partial_\alpha \tilde{\psi} - 
\frac{1}{\chi \omega_0^2} v^0 \tilde{\psi} v^\alpha \partial_\alpha \psi^* + \frac{1}{c} g v^0 (\psi^* \tilde{\varphi} - \tilde{\psi} \varphi^*)\right].
\eeq
This scalar product will be particularly important in the definition of the quantum states for the model at hand. 

The same scalar product can be obtained also as associated with the symplectic structure of the classical Hamiltonian equations. The starting point is still the complexified Lagrangian. One first defines
\beqnl
\pi_\varphi  &:=& \frac{\partial {\cal L}_c}{\partial \partial_0 \varphi}=\frac 12 \partial_0 \varphi,\\
\pi_\varphi^\ast  &:=& \frac{\partial {\cal L}_c}{\partial \partial_0 \varphi^\ast}=\frac 12 \partial_0 \varphi^\ast,\\
\pi_\psi &:=& \frac{\partial {\cal L}_c}{\partial \partial_0 \psi}=\frac{1}{2\chi \omega_0^2} v^0 v^\alpha 
\partial_\alpha \psi +\frac {g}{2c} v^0 \varphi,\\ 
\pi_\psi^\ast &:=& \frac{\partial {\cal L}_c}{\partial \partial_0 \psi^\ast}=\frac{1}{2\chi \omega_0^2} v^0 v^\alpha 
\partial_\alpha \psi^\ast +\frac {g}{2c}\varphi^\ast. 
\eeqnl
Then, introducing 
\beq
\Omega:=-i \left[
\begin{array}{cc}
0 & 1\cr
-1  & 0
\end{array} \right],
\eeq
where the square matrix represent the symplectic form of standard Hamiltonian classical mechanics,  
we can obtain the conserved scalar product as follows:
\beq
< \Psi_1, \Psi_2> := (\Psi_1, \Omega \Psi_2) = : \int dx \Psi_1^\ast \cdot \Omega \Psi_2,
\eeq
where $\cdot$ stays for the ordinary scalar product in ${\mathbb R}^4$, and 
\beq
\Psi:=\left(
\begin{array}{cccc}
\psi\cr
\varphi\cr
\pi_\psi\cr
\pi_\varphi
\end{array} \right).
\eeq

\subsection{Separation of variables in the comoving frame for $v=$const}

From the system (\ref{eq-psi},\ref{eq-varphi}) is not immediately evident that it is possible to 
separate variables in the comoving frame  for $v=$const. 
As to the dependence on transversal spatial variables $y,z$, it is evident that separation of variables occurs. 
In the case of the time variable $t$, it is simply necessary to write the system (\ref{eq-psi},\ref{eq-varphi}) 
as a first order system in $t$; in other words, we have to re-write the aforementioned system of equations of 
motion in Hamiltonian form. 
Then we obtain the following Hamiltonian form of the system:
\beq
\label{eq-hami-form}
i \partial_0 \left(\begin{array}{c}
\varphi\\
\psi \\
\pi_\varphi\\
\pi_\psi \\
\end{array}
\right)=
i \left[ \begin{array}{cccc}
0 & 0 & 1 & 0\cr
-\frac{g}{c} \chi \omega_0^2 &  -\frac{v^k}{v^0} \partial_k &  0 &  -\frac{\chi \omega_0^2}{v^0} \cr
(\nabla^2-\frac{g^2}{c^2}  \chi \omega_0^2 )
&
0
&
0
&
\frac{g}{c}\chi \omega_0^2\cr
0
&
-\frac{1}{\chi}
& 
0
&
-\frac{v^k}{v^0} \partial_k
\end{array}
\right]
\left(\begin{array}{c}
\varphi\\
\psi \\
\pi_\varphi\\
\pi_\psi \\
\end{array}
\right).
\eeq
In a more concise form, we can write the previous equation as follows:
\beq
i\partial_t \Psi = H \Psi,
\eeq
where $\Psi := (\varphi,\psi,\pi_\varphi,\pi_\psi,)^T$ and where $H$ is the matrix operator displayed in 
(\ref{eq-hami-form}). It is to be noted that $H$ is formally self-adjoint in the scalar product $<,>$ we have defined in the previous subsection. Indeed, it is not difficult to show that the hermitian conjugate of $H$, 
say $H_c=(\Omega H \Omega)^\dagger$, has the same form as the operator $H$.\\ 
Being $H$ independent on $t$, it follows that one can find stationary 
solutions in the form
\beq
\Psi = \exp (-i\omega t) F(x,y,z),
\eeq
and separation of variables in $t$ becomes evident. Notice that $\omega$ is a conserved quantity, 
and that this amounts to the following conservation in the lab frame
\beq
\omega_{lab}-v k_{lab, x} = {\mathrm{const}},
\eeq
which was previously obtained in a perturbative approach and now is confirmed in an exact 
model. Note that this conservation is simply associated with the conservation of energy 
in the comoving frame.

\subsection{Norm of the states and field quantization}

\begin{figure}[t] 
\includegraphics[angle=0,width=8cm]{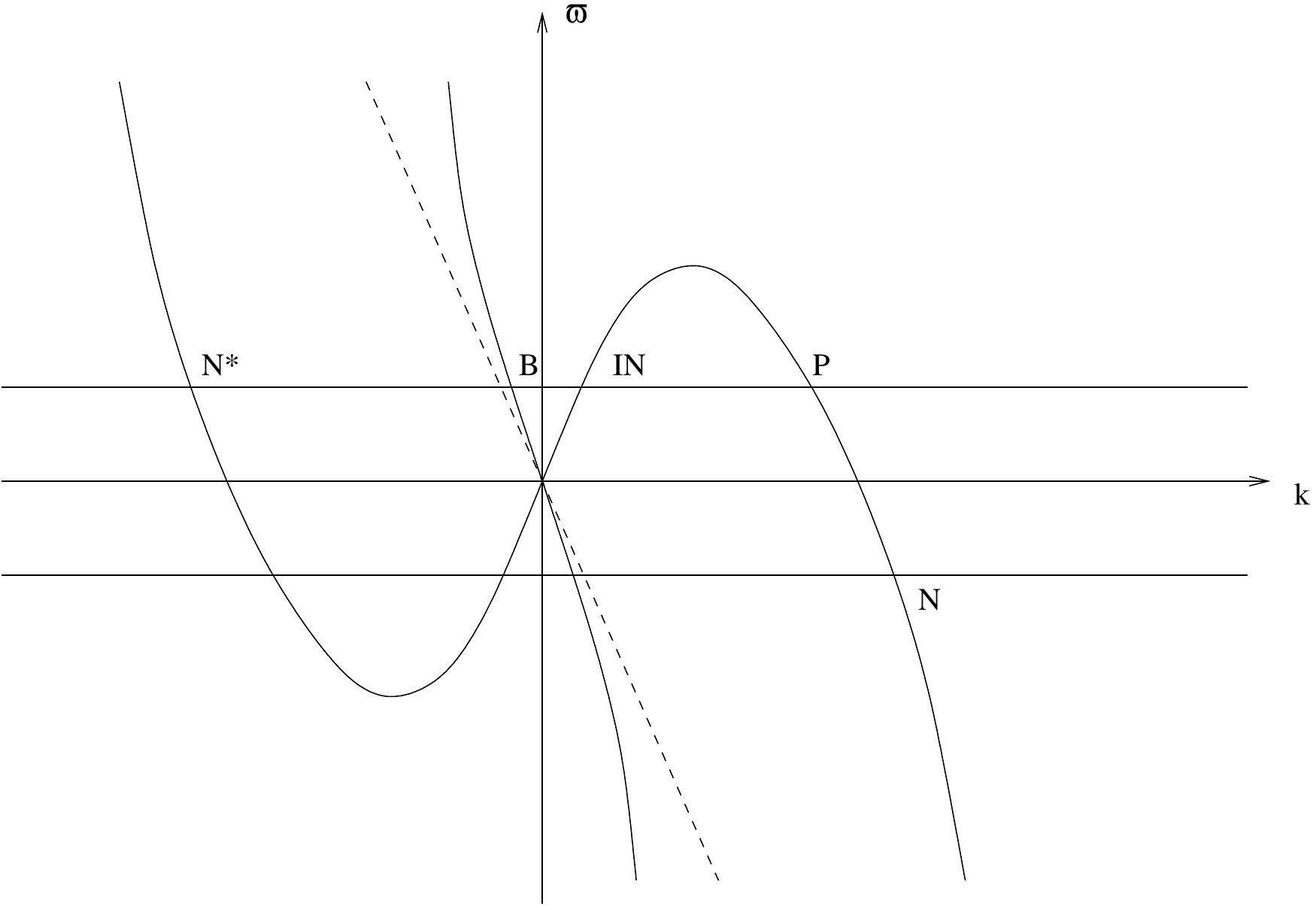}
	\caption{ \label{fig:fig1} Asymptotic dispersion relation for the Cauchy case in the comoving frame. The monotone branch is $G_+$, the non-monotone one is 
	$G_-$. The dashed line divides antiparticle states (below it) and particle ones (above it). Two lines at $\omega=$const and at $-\omega=$const are
	also drawn, and relevant states introduced in the text are explicitly  indicated.}
	\label{fig:asymptDR}
\end{figure}

The norm of the states is inherited by conserved scalar product:
\beq
||\Psi||^2:=<\Psi, \Psi> =  \frac{i}{2} \int_{\Sigma_t} \left[ \varphi^* \partial^0 \varphi - 
(\partial^0 \varphi^*) \varphi 
+ \frac{1}{\chi \omega_0^2} v^0 \psi^* v^\alpha \partial_\alpha \psi - 
\frac{1}{\chi \omega_0^2} v^0 \psi v^\alpha \partial_\alpha \psi^* + \frac{g}{c} v^0 (\psi^* \varphi - \psi \varphi^*)\right].
\eeq
Particle states correspond to positive norm states (which is a notion which remains invariant under Lorentz 
group), negative norm states correspond to antiparticles. This can be exemplified in the homogeneous case, 
i.e. in absence of the perturbation. Then, in the case of plane-wave solutions, it can be shown that 
\beq
||\Psi||^2:=< \Psi, \Psi> \propto \omega_{lab},
\eeq
so that $\sgn{||\Psi||^2}=\sgn{\omega_{lab}}$.\\
This results matches well what happens in the electromagnetic case, and shows that particles in the 
lab are defined by the condition $\omega_{lab}>0$. Then, in the comoving frame, we have that particle states 
are defined by the condition $\omega > -v k$. These conditions should represent a good indication for particle and antiparticle states also 
for the full problem. See Appendix \ref{scattering-asympt} for more details.  
This corroborates the common use of the asymptotic dispersion relation, to be indicated as (DR)-asymptotic 
in the following, in order to identify particle and antiparticle states. See figure \ref{fig:fig1}. For example, in the Cauchy approximation 
we have 
\beq
n(\omega_{lab})=n(0)+B \omega_{lab}^2+\delta n (x-vt),
\label{cau-n}
\eeq
where $n(0)$ is $n(\omega_{lab}=0)$, $B$ is constant, and the Kerr effect induces the right-moving perturbation $\delta n (x-vt)$;  
the (DR)-asymptotic displays three states on the same branch $G_-$ (cf. equation \eqref{c-branches}) with the same $\omega$ in the comoving frame: 
the positive group velocity particle state, to be called $IN$, the negative group velocity particle state, to be called $P$, 
and a negative group velocity antiparticle state $N^\ast$. It is also to be remarked that each state $(\omega, k)$ is conjugated to a state with  $(-\omega, -k)$ (and then there is e.g. a state $N$ which is the conjugate of  $N^\ast$).
See \cite{rubino-prl, rubino-sr}.  
There is also another branch $G_+$ (cf. equation \eqref{c-branches}), which for the same $\omega$ displays only a single negative group velocity particle state $B$. Only the former branch $G_-$ is involved in the Hawking process.\\

Quantization follows the general lines we have found for the electromagnetic case \cite{belcacciadalla-hopfield}, with the important 
simplification associated with the absence of constraints, which makes easier and standard the treatment of the 
topic. In particular, in the comoving frame (which is static), we have
\beq
\Psi (t,{\mathbf{x}})=
\sum \int \frac{d\omega}{2\pi} \frac{d{\mathbf{k}}}{(2\pi)^3} \frac{1}{N_\Psi} \delta (\textrm{dispersion relation})
\left( a(\omega,{\mathbf{k}}) U(\omega,{\mathbf{k}};t,{\mathbf{x}})+ 
 a^\dagger (\omega,{\mathbf{k}}) V(\omega,{\mathbf{k}};t,{\mathbf{x}}) \right),
\label{quant-field}
\eeq
where $N_\Psi$ stays for a normalization factor, $\delta (\textrm{dispersion relation})$ indicates that 
one considers solutions `on shell', and $U,V$ are positive norm solutions and negative norm solutions of 
the field equations respectively. To be more specific, we also point out that we are interested in the following fields: 
\beq
\Psi_{(IN)} (t,{\mathbf{x}}),
\eeq
which represents the field asymptotically in the past: as $t\to -\infty$, and the corresponding 
asymptotic field in the future: 
\beq
\Psi_{(OUT)} (t,{\mathbf{x}}).
\eeq
A single (rapid) transition from a region with a given set of parameters $\chi,g,\omega_0$ to a region with 
parameters  $\chi',g',\omega'_0$ is considered, and the above asymptotic field are interpolated by the field 
(\ref{quant-field}).\\
Notice that the above quantization, in a two-dimensional model, on shell, can give rise both to an $\omega$-representation, 
where only the integration in $d\omega$ is left, and also to a $k$-representation, where the dispersion relation 
is used to leave only the integral in $dk$. In a four-dimensional model, the latter choice is to be preferred.

In order to compute amplitudes for pair-creation, the strategy is the following. Let us expand in 
plane waves both the IN state and 
the $P,\; N,\; B$ states emerging from the scattering process. Then, let us consider 
\beq
|N|^2 := \frac{|J_x^N|}{|J_x^{IN}|},
\label{pair-creation}
\eeq
where $J_x$ stays for the conserved current (\ref{curr-j}), and the indexes N, IN indicate that one is considering 
the $N$-particle states and the $IN$-particle states respectively (and similar for $P$ and $B$). Indeed, $|N|^2$ above is the ratio between the outgoing 
flux along $x$ of negative energy particle states and the flux along $x$ of ingoing particle states, where an infinitesimal 
surface element orthogonal to $x$ has been simplified.  
$|N|^2$  coincides with the mean number per unit time and unit volume of created particles 
\cite{damour,nikishov,gitman}.  See also \cite{manogue}.            {\color{black} To this topic is devoted Appendix \ref{manley}, where 
scattering amplitudes are discussed in detail, and the relation between white hole scattering and black hole scattering 
is pointed out. Furthermore, a deduction of the generalized Manley-Rowe identitities (which are substantially 
Wronskian relations) is also given.   In the following section, we discuss particle creation and in particular its 
thermal character in our model in the Hawking case.}

\section{Analytical calculations and thermality}
\label{thermality}

We wish to set up an analytical approach allowing us to infer thermality for our model. Our 
reference works will be \cite{corley,cpf}. Indeed, we also try to 
identify a common mechanism for thermality, including our model, and the general fluid model discussed in 
\cite{cpf}, which is a generalization and a refinement of the seminal calculations by Corley \cite{corley}. 
The strategy is shared with the aforementioned calculations: WKB approximation is introduced at the 
level of the calculations for asymptotic states far from the horizon, and this expansion is matched 
with the near-horizon expansion, which is instead treated in Fourier space, in the approximation up to 
the linear order in $x$. It is not difficult to show that, under the hypothesis of not very strong 
gradient of the refractive index, one of the four asymptotic states (which belongs to the monotone branch of the 
dispersion relation) decouples and gives rise to a scattering phenomenon which is an almost negligible 
fraction of the dominant phenomenon represented by the Hawking effect. On the other hand, it is a consequence 
of the construction of the states in the near-horizon region that the states which match with the 
asymptotic states $P$ and $N$ (positive and negative norm states emerging from the scattering and lying 
on the non-monotone branch of the dispersion relation) are such that the ratio 
\beq
\frac{|N|^2}{|P|^2}=\exp \left( -\beta_h \hbar \omega\right),
\eeq
where $\beta_h$ is the black hole temperature, 
i.e. it leads to the standard thermal spectrum as far as the fourth state is negligibly coupled (which means 
that $|B|^2$ is negligible).\\

We start from the asymptotic expansion.

\subsection{WKB analysis}

We consider our equations of motion. Instead of working in full generality, we take into account only the possibility to get $g\mapsto g_0 + \delta g(x)$, where $g_0$ is constant. 
We could allow also a dependence on $x$ of the susceptibility $\chi \mapsto \chi_0+\delta \chi (x)$, as well as $\omega_0^2 \mapsto \omega_0^2+\delta \omega_0^2 (x)$,  where $\chi_0,\omega_0^2$ are constants, but 
this choice seems to represent an hindrance to a neat deduction of the Hawking effect, as we show in the 
following subsection. As a consequence, we avoid such a dependence, and justify this choice by recalling that our model 
is phenomenological: the travelling perturbation is introduced by means of inhomogeneities in the couplings 
of the model, which are chosen so that the refractive index displays the expected behaviour. What is important 
is the latter behaviour, and not the way the phenomenological model allows us to implement it. As a matter of 
facts, the model, as it is constructed, is able to reproduce a behaviour like $n(x_{lab}-v t_{lab},\omega_{lab})
=n_0 (\omega_{lab}) +\delta n ( x_{lab}-v t_{lab})$, with the travelling perturbation described as independent 
on the dispersion, only in the limit of negligible dispersion \cite{belgiorno-prd}, when the Cauchy approximation holds 
true. Otherwise, such phenomenological behaviour cannot be approached by means of the Hopfield model, and its extension 
to the nonlinear regime would be required.\\

All considerations above in any case do not affect the zeroth order WKB approximation, which is obtained as follows. 
As in \cite{corley,cpf}, we have $\partial_t \mapsto i \omega$. 
We adopt the 
following expansion:
\beq
\left(
\begin{array}{c}
\varphi\cr
\psi
\end{array} 
\right) = 
\sum_{k\geq 0} \left(\frac{\hbar}{i}\right)^k \exp (i\frac{\omega}{\hbar} t + i \frac{S(x)}{\hbar})
\left(
\begin{array}{c}
A_k\cr
B_k
\end{array} 
\right) .
\eeq
Then, {\color{black} defining $p=\partial_x S$,  we get the following zeroth order matrix by expanding in powers of $\hbar$:
\beq
M_{(0)}=\left[
\begin{array}{cc}
-\frac{\omega^2}{c^2}+p^2 & -i \frac{g}{c} \gamma (\omega+v p)\\
 i \frac{g}{c} \gamma (\omega+v p ) & \frac{1}{\chi \omega_0^2} (\omega_0^2 - \gamma^2 (\omega+v p)^2)
\end{array}
\right].
\eeq
As to the zeroth order, the mandatory vanishing of the determinant of $M_{(0)}$ amounts to 
\beq
c^2 p^2 -\omega^2 = \frac{g^2 \gamma^2 \chi (\omega+v p)^2}{1-\frac{\gamma^2 (\omega+v p)^2}{\omega_0^2}},
\label{det-m}
\eeq
which is  equivalent to the usual dispersion relation in the lab, apart for the dependence on space of the inhomogeneous 
terms. We obtain a polynomial of degree 4 in $p$, to which we can associate four solutions, as expected}.\\  
Due to the intrinsic nature of our model, it makes sense to adopt the Cauchy approximation. As it is easy to 
understand, at this level there is no substantial difference between solutions one obtains as above and the solutions 
one can obtain from the two branches of the Cauchy approximation in a phenomenological model where the 
eikonal approximation is assumed. We assume that 
\beq
p(x) = p_0 (x) + \epsilon p_1 (x) +\ldots,
\eeq
i.e. we assume that $p(x)$ is developable in series. 
Then, we consider the following perturbative solution ansatz, which is equivalent 
to the one adopted in \cite{cpf,corley}: 
\beqnl
p(x)&=&p_0 (x)+\epsilon p_1 (x),\\
\omega &=& \epsilon \omega_1,
\eeqnl
which is equivalent to the ansatz $\omega \ll p v$. The two branches in the Cauchy approximation 
are 
\beq
G_\pm := \gamma (\omega+ v p) (B \gamma^2 (\omega+ v p)^2 + n(x))  \pm \frac{c}{v}\left(\gamma (\omega +v p)
-\frac{\omega}{\gamma}\right) =0,
\label{c-branches}
\eeq
where $G_+=0$ corresponds to the monotone branch. From $G_-=0$, in the above perturbative ansatz, 
we obtain three solutions: 
\beqnl
p_\pm &=& \pm \frac{1}{\sqrt{B}} \frac{1}{\gamma v} \sqrt{\frac{c}{v}-n}
           -\frac{\omega}{v}-\frac{1}{2} \frac{c \omega}{\gamma^2 v^2} \frac{1}{\frac{c}{v}-n},\\
p_{+s} &=& -\frac{\omega}{v}+ \frac{c \omega}{\gamma^2 v^2} \frac{1}{\frac{c}{v}-n},
\eeqnl
where we have purposefully introduced a notation which resembles the one in \cite{corley}, 
and indeed the modes we have found correspond to the modes identified therein. It can be noticed that asymptotically 
the norm of $p_+,p_{+s}$ is positive, being $\omega+ v p>0$, whereas the norm of $p_-$ is negative. 
Moreover, $p_\pm$ are short wavelength states, and  $p_{+s}$ is a long wavelength one. A fourth real solution 
emerges from $G_+=0$, and is the one which decouples from the spectrum in the approximation where a not 
too strong gradient in the refractive index occurs. The above solutions have to be matched with the 
ones which will be obtained in the near horizon region. With this aim, we investigate also the amplitude part of the WKB solutions. 
{\color{black} We can obtain the same results from equation (\ref{det-m}). We are also interested in the amplitudes 
associated with the zeroth order equation, which leaves undetermined one solution (due to the fact that $M_{(0)}$, which is a  
rank 1 matrix, is to be considered `on shell', i.e. 
for $p$ corresponding to a root of  $ \det M_{(0)}=0$). We need to look for the first order equation, which is 
\beq
M_{(1)} 
\left(
\begin{array}{c}
A_0\cr
B_0
\end{array} 
\right)
+ M_{(0)} 
\left(
\begin{array}{c}
A_1\cr
B_1
\end{array} 
\right)=0.
\label{eq-m1}
\eeq
We have 
\beq
M_{(1)}=\left[
\begin{array}{cc}
-i (\partial_x p) -2 i  p \partial_x &  -\frac{1}{c} \gamma v g \partial_x\\
 \frac{1}{c} \gamma v (g \partial_x+( \partial_x g)) & i \frac{\gamma^2 v}{\chi \omega_0^2} ((\partial_x s) +2 s \partial_x )
 \end{array}
\right],
\eeq
where $p$ coincides with one of the roots of $\det M_{(0)}=0$ (at the leading order) and we introduced:
\beq \label{eq:s}
s=\gamma (\omega + v p).
\eeq
Following the theory of the multicomponent WKB (see e.g. \cite{ehlers,Fedoryuk}),  in order to find out 
the zeroth order amplitude, we proceed as follows. We first express the zeroth order solution as $r_0 a_0$, 
where $r_0$ is a right eigenvector of $M_{(0)}$ relative to the zero eigenvalue, and $a_0$ is the unknown amplitude.  
Then we project (\ref{eq-m1}) on the left eigenvector $l_0$  of $M_{(0)}$ relative to the zero eigenvalue, obtaining 
then a single equation $l_0 M_{(1)} r_0 a_0=0$ for the amplitude. A rather straightforward calculation 
gives us, in the limit of negligible $\omega$, and for the leading order solutions $p_\pm$ ($M_{(1)}|_{p_\pm}$ is considered) 
\beqnl
&& A_0 \propto \frac{1}{x^{3/4}},\\
&& B_0 \propto \frac{1}{x^{1/4}},
\eeqnl
in the near horizon region where the WKB approximation 
still holds, in the so called linear region of \cite{cpf}, but not too near the horizon, where it breaks down. 
This behavior is also expected from the near horizon expansion (see below equation (\ref{stapm}))}.

\subsection{Near horizon region analysis}
\label{thermality-near}

For the region near the horizon $x_+$ which is such that $n(x_+)=c/v$, we proceed again by adopting 
the same ansatz as in \cite{corley,cpf}. Our starting point is to Fourier transform the equations of 
motion, by keeping into account that $x\mapsto i \partial_p$. The trick is a linearization 
in the neighborhood of the horizon $x=0$: 
$n(x) \sim \frac{c}{v}+\kappa x \mapsto \frac{c}{v}+\kappa i \partial_p$.

\begin{figure}[t]
\includegraphics[angle=0,width=8cm]{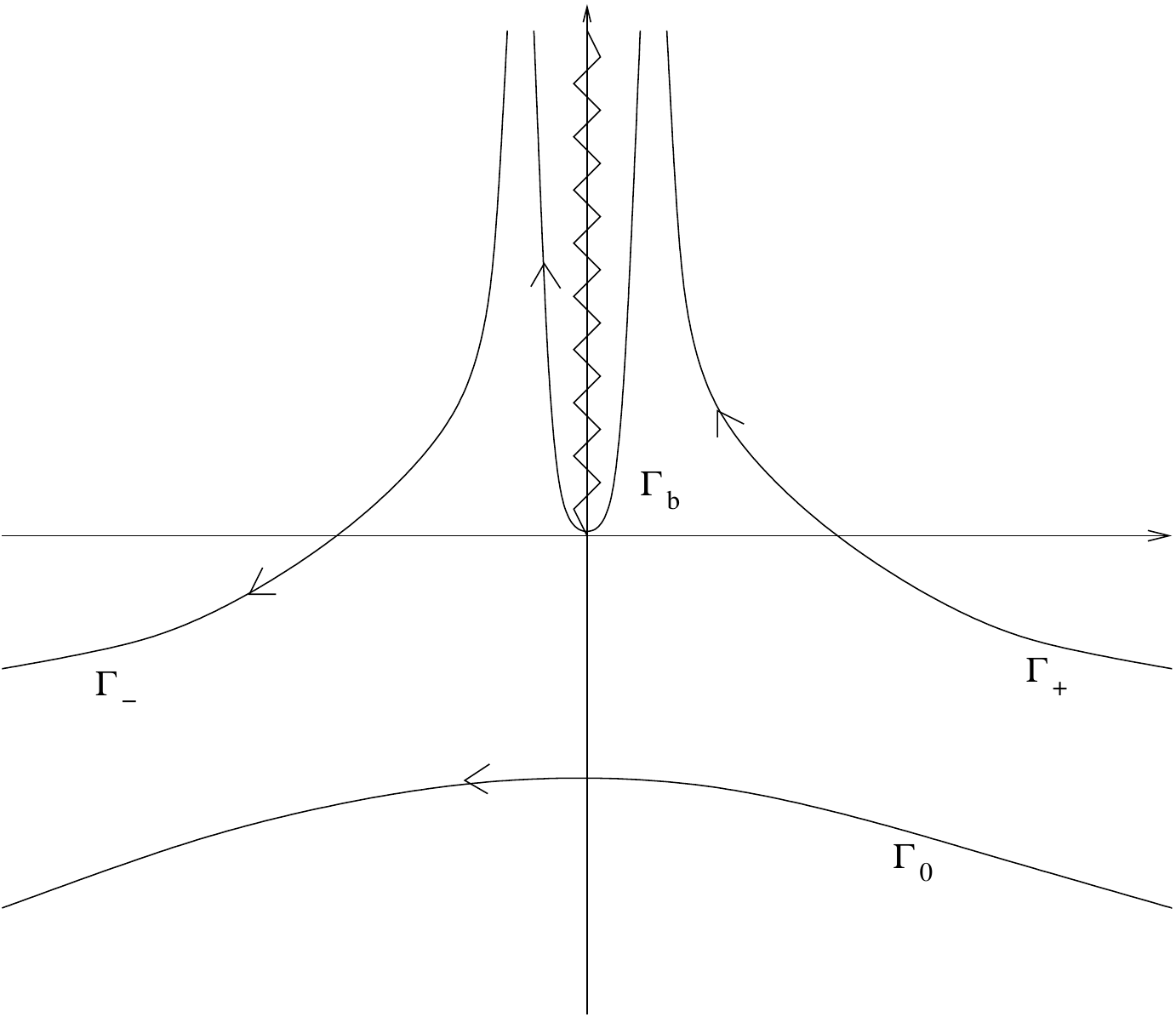}
	\caption{ \label{fig:fig2} Circuits in the complex $s$-plane defining the relevant modes as described in the text. }
\end{figure}

In the following analysis, our guide is 
represented by the presence of a suitable branch cut 
in Fourier space which should allow us to pinpoint the origin of thermality. 
At first, we proceed by looking for a Fuchsian singularity as the one occurring  at $p=0$ also in \cite{cpf,corley}. 
Our ansatz is that, given the universality in the Hawking effect in the nondispersive case, 
one should be able to find out a common origin of thermally also in the dispersive case. We show in the 
following that it is indeed possible to implement thermality according to our ansatz. 
As both our model and the ones in \cite{corley,cpf} 
characterize a huge class of analogous systems, we conjecture that such a behaviour 
is a distinctive signal for the Hawking effect in analogous systems. Indeed, provided 
that a suitable matching exists with WKB solutions in the asymptotic region, the 
calculation of thermality can be associated with a (hyper-)local behaviour, in the sense that, 
once the aforementioned matching allows to find a relation between asymptotic WKB 
wavefunctions and local functions near the horizon, thermality can be calculated 
even locally. It is sufficient to compare $|N|^2/|P|^2$ near the horizon to get the 
right behaviour. \\
The aforementioned ansatz 
is interesting, in the sense that 
it is very selective on the way the phenomenological model should be related to the behaviour of microscopical parameters 
appearing in the Hopfield model. Indeed, in the latter one, we could allow the following dependence on $x$ variable in the comoving frame of the dielectric 
perturbation: $\chi \mapsto \chi_0 +\delta \chi (x), \omega_0^2 \mapsto \omega_0^2+\delta \omega_0^2 (x), g\mapsto g_0+\delta g (x)$. The worst 
behaviour is obtained by the presence of a dependence on  $\delta \chi (x)$, as it does not allow any Fuchsian structure and makes as hard as possible 
the problem of identifying the source of thermality. Even by considering only a first order equation arising from equation 
(\ref{eqgfou}) (see below), it is not possible to identify a mechanism for thermality. 
So our choice is to factor out this behaviour. There is no problem, in our view, in this choice, as 
in any case the Hopfield model we are trying to adapt in order to match the phenomenological behavior is only an approximate model which should 
be more correctly described by a nonlinear version of quantum electrodynamics (as the dielectric perturbation arises from nonlinearities of the 
dielectric medium). Then we consider the correction $\delta \omega_0^2 (x)$, which is interesting (indeed, it could be as well considered as arising 
from a linearization of the nonlinear polarization term $P^4$ one could add to the Lagrangian). Still, even if a Fuchsian structure can be identified, the 
roots of the indicial equation depend on $\omega_0^2$ instead than on $\omega^2$, so that again thermality is not extracted from the model. 
The last possibility, which is to some extent an unexpected variant of the model, because it amounts to a variation of the plasma frequency, but at constant susceptibility $\chi$, is 
at the very least satisfactory. Indeed, it not only provides us a good Fuchsian behavior with thermality, but it is also a good model at the level of a first 
order analysis.\\ 
We also symmetrize the last term in the Lagrangian (\ref{lagrangian}): 
$\frac{g}{c} (v^\alpha \partial_\alpha \psi) \varphi \mapsto \frac{g}{2c} (v^\alpha \partial_\alpha \psi) \varphi -\frac{g}{2c} (v^\alpha \partial_\alpha \varphi) \psi$. 
From the equations of motion we obtain 
\beq
\tilde{\varphi} (p) = \frac{1}{p^2-\frac{\omega^2}{c^2}} \frac{1}{c} \left[ \frac{1}{2} \alpha \gamma v + i g_0 \gamma (\omega + v p) -\alpha \partial_p 
 \gamma (\omega + v p) \right] \tilde{\psi} (p),
\label{equaphi}
 \eeq
where
\beq
g(x)=g_0+\alpha x \mapsto g_0+i \alpha \partial_p,
\label{eqgfou}
\eeq
and 
\beq
\frac{1}{\chi} \left( 1-\frac{\gamma^2 (\omega + v p) ^2}{\omega_0^2}\right) \tilde{\psi} (p)+\frac{1}{c} \left(\frac{1}{2} \alpha \gamma v + i g_0 \gamma (\omega + v p)  
-\alpha \partial_p \gamma (\omega + v p) \right) \tilde{\varphi} (p) = 0.
\label{equapsi}
\eeq
Equations (\ref{equaphi}, \ref{equapsi}) lead in a natural way to a second order differential equation, which is better described in terms of the 
variable $s$ and  which displays a Fuchsian singularity  for $s=0$
\beq
\left[\partial_s^2 +\frac{2}{s} \partial_s + \left( \frac{1}{4}+\frac{c^2 \omega^2}{\gamma^4 v^4 \chi \alpha^2}\right) \frac{1}{s^2}\right] \tilde{\psi} (s)=0,
\eeq
whose indicial roots are
\beq
\alpha_\pm= -\frac{1}{2}\pm i \frac{c \omega}{\gamma^2 v^2 \sqrt{\chi} \alpha}.
\eeq
Then, near $s=0$, the solutions behave as 
\beq
\tilde{\psi} (s) \sim s^{-\frac{1}{2} \pm i \frac{c \omega}{\gamma^2 v^2 \sqrt{\chi} \alpha}}.
\eeq
We note that a similar 
Fuchsian singularity appears 
also in \cite{cpf,corley}. As both the models 
characterize a huge class of analogous systems, we conjecture that such a behaviour 
is a distinctive signal for the Hawking effect in analogous systems.
\\
By comparing with the phenomenological dispersion relation $G_+ G_-=0$  in the Cauchy approximation (the latter approximation  is the only one to be expected to provide a 
sensible match between the Hopfield model and the zeroth order assumption of non-dependence on $\omega$ of the dielectric 
perturbation $\delta n (x)$), we find 
\beqnl
g_0 &=& \sqrt{\frac{n(0)^2 -1}{\chi}},\\
\delta g &=& n(0)\sqrt{\frac{1}{(n(0)^2 -1)\chi}}\delta n,\\
\omega_0^2 &=&\frac{n(0)^2 -1}{2 n(0) B}.
\eeqnl
We assume $\chi>0,g_0>0$. 
In particular, we stress the following relation:
\beq
\alpha:=g'= n(0)\sqrt{\frac{1}{(n(0)^2 -1)\chi}} n',
\label{eqalfa}
\eeq
where the prime indicates the $x$-derivative and the above identity is considered at $x=0$. Note that $n'=\kappa$, which is not yet the 
surface gravity (because also of the sign to be taken into account).\\
Even if an almost satisfactory behaviour can be identified, there is a problem: it is not possible to find out a solution for (\ref{equapsi}) in 
explicit form, so it is not clear how to prescribe the behaviour of the solutions. As a consequence, in the spirit of our above considerations, we turn to 
a first order analysis. This amounts to consider (\ref{equapsi}), where $\tilde{\varphi}$ is given by (\ref{equaphi}), and to neglect terms $\propto \alpha^2$.\\ 
{\color{black} We assume that $\alpha \gamma v \ll s$ in order to properly deal with the limit as $\omega\to 0$}.\\
The first order equation one obtains admits the following solution:
\beq
\tilde{\psi}(s) = \sqrt{c^2 (s-\gamma \omega)^2 -\gamma^2 v^2 \omega^2} \frac{1}{s}  \exp (i q(s)),
\eeq
where the phase $q(s)$ is 
\beq
q(s):= \frac{c^2 s^3}{6\alpha \chi g_0 \gamma^3 v^3 \omega_0^2}+ \frac{c}{v} \frac{c \omega}{\alpha \chi g_0 \gamma^2 v^2}\log (s) +w(s),
\eeq
where
\beq
w(s):=
\frac{g_0 s}{2 \alpha \gamma v}+\frac{c^2 \omega^2}{2 s \alpha \chi g_0 \gamma^3 v^3}-\frac{c^2 s}{2 \alpha \chi g_0 \gamma^3 v^3}
-\frac{c^2 s^2 \omega}{2\alpha \chi g_0 \gamma^2 v^3 \omega_0^2}+
\frac{c^2 s \omega^2}{2\alpha \chi g_0 \gamma v^3 \omega_0^2}-\frac{s \omega^2}{2 \alpha \chi g_0 \gamma v \omega_0^2}.
\eeq
We aim to write the field $\psi (t,x)$ as follows:
\beq
\psi (t,x)=e^{i\omega t} \int_\Gamma \; dp\; \frac{1}{\sqrt{2\pi}} \tilde{\psi} (p) e^{i p x}=e^{i\omega t}  e^{-i \frac{\omega}{v} x}\frac{1}{\gamma v}\int_\Gamma \; ds\; \frac{1}{\sqrt{2\pi}} \tilde{\psi} (s) e^{i s \frac{1}{\gamma v} x},
\eeq
for a suitable path $\Gamma$ in the complex plane.
\\
Our ansatz is the following: the contribution $w(s)$ is not relevant in the saddle point approximation, in such a way that, keeping into 
account the contribution of the Fourier transform,  
the saddle points are determined by 
\beq
\frac{d}{ds} \left(\frac{s}{\gamma v} x + \frac{c^2 s^3}{6\alpha \chi g_0 \gamma^3 v^3 \omega_0^2}\right)=0.
\label{saddle-p}
\eeq
In other terms, the relevant saddle points are the ones of the function 
\beq
h(s):=\frac{s}{\gamma v} x + \frac{c^2 s^3}{6\alpha \chi g_0 \gamma^3 v^3 \omega_0^2}.
\eeq
 According to the saddle point 
approximation we have 
\beq
\int_\Gamma \; ds\;  F(s) \exp(i h(s)) \simeq \sum_{k} F(s_k) \exp(i h (s_k) ) \frac{1}{\sqrt{-i \frac{d^2 h}{ds^2} (s_k) }},
\eeq
where the sum is extended to the saddle points. From (\ref{saddle-p})
We obtain 
\beq
\frac{c^2 s^2}{2\alpha \chi g_0 \gamma^2 v^2 \omega_0^2}+x=0,
\eeq
i.e. 
\beq
s_\pm = \pm \sqrt{\sign (x) |x|} \sqrt{2 \chi |\alpha| g_0} \gamma v \omega_0 \frac{1}{c}:=\pm\eta|x|^{1/2}, 
\eeq
where we have taken into account that $\alpha<0$. 
In particular, we get 
\beqnl
&&a)\ x>0 \Rightarrow s_\pm = \pm  (\sqrt{2 \chi |\alpha| g_0} \gamma v \omega_0 \frac{1}{c}) x^{1/2},\\
&&b)\ x<0 \Rightarrow s_\pm = \pm  i(\sqrt{2 \chi |\alpha| g_0} \gamma v \omega_0 \frac{1}{c}) |x|^{1/2},
\eeqnl
where the case $(a)$ refers to the outer region and $(b)$ to the inner one. 
The large parameter in the saddle point approximation is 
\beq
\sqrt{2 |\alpha| g_0 \chi \omega_0^2 } = \sqrt{\frac{n(0)^2 -1}{B} |\kappa|} \propto \sqrt{\frac{|\kappa|}{B}} ;
\eeq
the above parameter is manifestly large due to the smallness of $B$. \\
Due to the strong analogy with \cite{corley,cpf} case, we can adopt the same choice for 
circuits in the complex plane (the only difference being in using variable $s$ instead of variable $p$).  
See Fig. \ref{fig:fig2}.
Forbidden sectors in the complex $s$-plane are formally the same, albeit $s$ cannot be too high 
because in our approximation $\omega_0$ acts as ultraviolet cut-off $s \ll \omega_0$. \\
The case (b) above provides us modes with exponentially increasing/decreasing behaviour. The 
growing mode is not physically allowed, whereas the decreasing one is assumed to be defined 
along the curve $\Gamma_0$ passing through $-i (\sqrt{2 \chi |\alpha| g_0} \gamma v \omega_0 \frac{1}{c}) x^{1/2}$ 
and with asymptotes in the allowed regions in the lower $s$-plane. As to the branch point $s=0$, 
we choose the associated branch cut along the positive $\Im(s)$ axis. The decreasing mode, in the 
saddle point approximation, is
\beq
\psi_0 (x) \sim C_0 \exp (-\frac{4}{3} |x|^{3/2} \frac{\eta}{\gamma v}) f(x) \frac{1}{|x|^{1/4}},
\label{stazero}
\eeq
where $C_0$ is a normalization constant.
The function $f(x)$ is associated with the remaining dependence on $x$ (it depends also on $\eta$).\\  
Modes with high momentum correspond to curves $\Gamma_\pm$, passing through 
$\pm (\sqrt{2 \chi |\alpha| g_0} \gamma v \omega_0 \frac{1}{c}) |x|^{1/2}$, ending in the allowed region 
and flowing asymptotically near the branch cut (see figure). $\Gamma_{-}$ flows from the allowed left 
region to the allowed region having the positive imaginary $\Im (s)$-axis as asymptote. 
$\Gamma_{+}$ arises from the left of the aforementioned branch cut (asymptote) and ends in the 
right allowed region. We get 
\beq
\psi_\pm (x) \sim C_+ \exp (i r_\pm (x)) \frac{1}{|x|^{1/4}} \exp (\pm \pi  \omega 
\frac{c}{v} \frac{1}{n(0)} \frac{c}{|\kappa| \gamma^2 v^2}),
\label{stapm}
\eeq
where $C_+$ is a (common) normalization constant and $r_\pm$ indicates a pure phase function. The 
real `thermal' characterization arises from the real exponential term, with exponent proportional 
to $\omega$. In particular, thermality is related to the contribution 
\beq
-\frac{c}{v} \frac{c \omega}{|\alpha| \chi g_0 \gamma^2 v^2}\log (s) 
\eeq
in the phase factor, by taking into account that  
\beq
\log (s\pm i \epsilon) \rightarrow \log |s| \pm i \pi \quad \quad \textrm{as}\ \epsilon\to 0.
\eeq
One could wonder if this corresponds to a good temperature for our model, matching the one for the 
nondispersive model. The answer is positive, indeed 
\beq
\beta_H=\frac{2\pi k_b}{\hbar} \frac{c}{v} \frac{c}{|\alpha| \chi g_0 \gamma^2 v^2}=\frac{c}{v} \frac{1}{n(0)} \left( \frac{2\pi k_b}{\hbar} \frac{c}{|\kappa| \gamma^2 v^2} \right),
\label{temp-bh}
\eeq
where the term in brackets is immediately realized to have the same form of the inverse of the nondispersive temperature. As it stands, the factor 
$\frac{c}{v} \frac{1}{n(0)} $ is substantially equal to one, because the horizon condition $n(0) + \delta n(x)=c/v$ implies, due to 
the smallness of $\delta n$, that $n(0)\sim c/v$. Still, it is true that it is not an exact result. 
The model is imperfect, which is to be expected.\\
It is interesting to point out that pair creation, in a thermal way, comes out because of the 
circuits suitably running around the branch point $s=0$, which is also the threshold 
for defining particle ($s>0$) and antiparticle ($s<0$) states.\\  
There is a fourth circuit $\Gamma_b$ which 
is running around the branch cut. Its behavior can be inferred by analogy 
with \cite{corley,cpf}.\\

Near the horizon, for $\omega \to 0$ (cf. \cite{corley,cpf}),  
the contribution to the amplitude arising from the saddle points is order of $1/\sqrt{s}$, i.e. 
$x^{-1/4}$, which matches the WKB result in the same order of approximation. Of course, one has to keep into account if 
$\sign (x)$ is positive or negative (i.e. if the mode is in the inner part or in the outer part of the horizon). In the 
inner region, there is an evident decreasing exponential factor which tames the modes, as in \cite{corley,cpf}.\\ 
{\color{black} As to the behavior of the field $\varphi (x)$, by taking into account that $\alpha \gamma v \ll s$, 
one obtains $\varphi_\pm \propto x^{-3/4}$, again in agreement with the WKB calculations.}\\

In particular, we are interested in the large $p$ modes, which amount to modes $p_\pm$ in the WKB approximation. 
These modes are involved with $\Gamma_\pm$ which give rise to the well-known thermality factor, with temperature 
\beq
T_H \simeq   \frac{\hbar}{2 \pi c k_b}  |\kappa| \gamma^2 v^2. 
\label{temp-teor}
\eeq
We could proceed as in Corley and CPF for getting this result, or we can adopt a simpler ansatz, which is grounded 
on the nature of modes $p_\pm$ to be associated with positive and negative norm states with an `ultraviolet' 
momentum. They correspond to $P$ and $N$ modes respectively, and, as far as the fourth mode $B$ is 
weakly coupled to the scattering, thermality can be inferred simply by the behavior of the ratio 
\beq
\frac{|N|^2}{|P|^2}=\exp \left(-\frac{\hbar \omega}{k_b T_H}\right), 
\eeq
where $T_H$ is given by (\ref{temp-teor}).

\section{Expansion point: which kind of horizon do we need?}
\label{horizons}

We have assumed, up to now, to work with the nondispersive (geometrical) horizon NDH 
as expansion point where the near horizon analysis 
is carried out. As a matter of facts, this choice is not the only one. Indeed, we could as well choose two other horizons: the group 
horizon GH and the phase horizon PH, 
which are defined respectively as the blocking horizon for propagating waves at a given frequency in 
the comoving frame (the point where the group velocity of the wave packet vanishes), 
and the locus where the phase velocity of the waves composing the wave packet vanishes. From a physical point of view, at first 
sight, there is no doubt that GH is more attractive and meaningful than the PH, and should play the role of NDH in the 
dispersive case, as it results in the optical case in \cite{petev-prl}, where GH is referred to as `blocking horizon'. Pure thermality 
of the spectrum is supposed to rely on the presence of a GH. 
Still, we wish to consider both concepts, as they could play a role which is not yet made as evident from our 
previous calculations.

In our analysis in subsection \ref{thermality-near} {\color{black} the horizon surface 
is at a generic locus simply indicated as `horizon'} and shifted to $x=0$, where only at the end of calculations our microscopic 
parameters are transformed into the macroscopic ones. Equation (\ref{temp-bh}) remains unaltered, and changes 
occurs only at the level of the subsequent approximation for  the factor 
$\frac{c}{v} \frac{1}{n(0)}$. It is still true that it is substantially equal to one, 
because  $n(0)\sim c/v$. Corrections in the temperature appear, depending on $\omega$ in the GH case, and on $k_0$ in the 
PH case. In particular, one finds
\beqnl
\frac{ \frac{c}{v} }{ \frac{c}{v}-n_{PH} } &=& \frac{1}{1-\frac{v}{c} z k_0^2} \sim 1+\frac{v}{c} z k_0^2,\\
\frac{ \frac{c}{v} }{ \frac{c}{v}-n_{GH} } &=& \frac{1}{1-\frac{v}{c} \zeta_B \omega^{2/3}} \sim 1+\frac{v}{c} \zeta_B \omega^{2/3},
\eeqnl
provided that in the former case $\frac{v}{c} z k_0^2\ll 1$ holds true, as well as $\frac{v}{c} \zeta_B \omega^{2/3}\ll 1$ in the latter case and $\zeta_B$ is given by equation \eqref{gh-cau-explicit}. 
No modifications in the saddle point approximation and in the choice of the paths appears, as quantities appearing in the 
calculations, like $\kappa$, do not depend on $s$ (they can depend on $\omega$ or on $k_0$, which are fixed parameters).\\
{\color{black} As a matter of facts, we have also to point out that in fluid models the above distinction between 
different kind of horizons, and in particular between nondispersive horizon and group horizon, is {\sl a posteriori} irrelevant, in the sense that 
conditions can be given such that the wave function of the modes is not able to distinguish between the aforementioned horizons, and also any 
correction to thermality is washed out. This is the main contribution of Ref. \cite{coutant-broad}. Due to 
several analogies between the framework discussed therein and our one, our hypothesis is that a similar conclusion is reasonable also in our case, despite the formal dependence we found above on the 
expansion point. We do not delve into this question 
herein.}

\section{Multi-resonances case for the Hopfield model}
\label{multibranches}

We discuss shortly in this section how to modify the model in order to take into account several resonances 
$\omega_{01}<\omega_{02}<\ldots<\omega_{0N}$.   
We introduce
\beq
{\cal L} = \frac{1}{2} (\partial_\mu \varphi)(\partial^\mu \varphi)+\sum_{i=1}^N 
\left\{\frac{1}{2\omega_{0i}^2 \chi_i} \left[ (v^\alpha \partial_\alpha \psi_i)^2 - \omega_{0i}^2 \psi_i^2 \right] 
+ \frac{g_i}{c} (v^\alpha \partial_\alpha \psi_i) \varphi\right\},
\eeq
where $N$ polarization fields $\psi_i$ appear, as well as $N$ a priori different $\omega_{0i},\chi_i,g_i$. 
As a consequence, in the homogeneous case we get
\beq
n^2 = 1+ \sum_{i=1}^N \frac{g_i^2 \chi_i}{1-\frac{s^2}{\omega_{0i}^2}}.
\eeq
It is rather simple to manage expressions like 
\beq
\frac{1}{1-\frac{s^2}{\omega_{0i}^2}}= \left\{
\begin{array}{lr}
\sum_{k=0}^\infty \left(\frac{s^2}{\omega_{0i}^2}\right)^k &  \mathrm{for}\ s<\omega_{0i},\\
-\frac{\omega_{0i}^2}{s^2}\sum_{k=0}^\infty \left(\frac{\omega_{0i}^2}{s^2}\right)^k &  \mathrm{for}\ s>\omega_{0i}.
\end{array}
\right.
\eeq
This allows to find expressions to the desired order of approximation for any value of $s$. The only 
warning is represented by the regions in a strict neighbourhood of the resonances themselves. Such regions 
cannot be described through the Hopfield model in the present simplified version, because there a 
large absorption by the dielectric medium, whose dissipative aspects require a special care, takes place. We simply 
limit ourselves to neglect these regions, deserving their study to more complete models. As to the 
Hawking effect in dielectric media, we point out that the near horizon analysis does not introduces 
any new interesting features: indeed, the only Fuchsian term in a second order expansion, or the only 
logarithmic branch point in a first order expansion still occur at $s=0$. E.g. with reference to the former 
expansion, and limiting ourselves to a two-resonance ($N=2$) model, {\color{black} where the two polarization fields 
$\tilde{\psi}_1,\tilde{\psi}_2$ are re-expressed in terms of $\tilde{\varphi}$ by using the equations of motion,} one finds 
\beq
\left[\partial_s^2 +\frac{2}{s} \partial_s + \left( \frac{1}{4}+\frac{c^2 \omega^2}{\gamma^4 v^4 (\chi_1 \alpha_1^2+\chi_2 \alpha_2^2)}
\right) \frac{1}{s^2}\right] \tilde{\varphi} (s)=0,
\eeq
where $\alpha_i$ amounts to the derivative of $g_i$, in a straightforward generalization of the single-branch model. 
Microscopic parameters are related to macroscopic ones as follows in the region $s<\omega_{01}$:
\beqnl
&&g_1^2 \chi_1+g_2^2 \chi_2 = n(0)^2 -1,\\
&&g_1 \chi_1 \delta g_1 +g_2 \chi_2 \delta g_2 = n(0)\delta n,\\
&&g_1^2 \chi_1 \frac{1}{\omega_{01}^2}+g_2^2 \chi_2\frac{1}{\omega_{02}^2} =B\sqrt{n(0)^2-1}.
\eeqnl
We also get the following relation:
\beq
n' n(0) = g_1 \chi_1 \alpha_1 +g_2 \chi_2 \alpha_2.
\label{temp-multi}
\eeq
A first order expansion in $\alpha_1,\alpha_2$, allows to recover a phase term related to thermality of the form 
\beq
i \frac{c^2 \omega}{( g_1 \chi_1 \alpha_1 +g_2 \chi_2 \alpha_2) \gamma^2 v^3}  \log \left( s\right); 
\eeq
it is easy to show that, thanks to (\ref{temp-multi}), one is able to recover again (\ref{temp-bh}).  
It is evident that the model is more involved, and is remarkable that there is the possibility to 
have more complex scattering processes, as it is evident also by inspection of the dispersion relation.
\begin{figure}[t]
\includegraphics[angle=0,width=10cm]{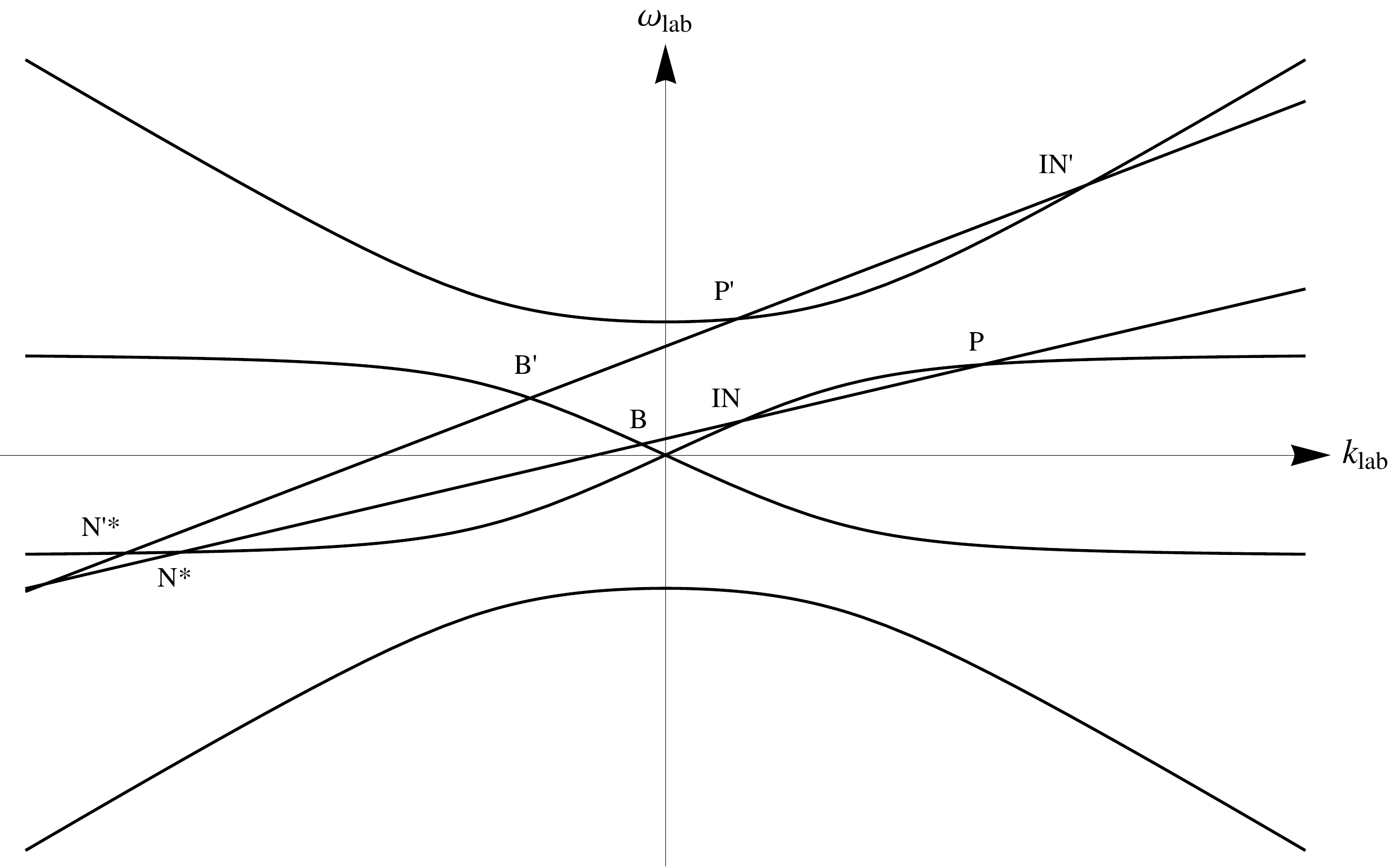}
	\caption{ \label{fig:fig3} Asymptotic dispersion relation for the Sellmaier dispersion relation of a diamond-like material (qualitative plot). 
The lab frame is adopted. Lines of constant $\omega$ are represented by straight lines.}
\end{figure}
In particular, there are processes which involve both the lower branch and the upper one, and which 
are candidated to be Hawking-like. For example, a process $IN'\to P'+N'^{\ast}$ is possible, where $N'^{\ast}$ 
belongs to the lower branch and $P',IN'$ to the upper one (see Fig. \ref{fig:fig3}). We shall consider details in a future publication, 
and limit ourselves to observe that in the latter case $IN'$ is a higher energy and momentum state than 
$P'$, and then, whichever nature one would be able to attribute to the aforementioned scattering, 
what is sure is that vacuum emission still will be peaked at the lower branch emission process we have 
analyzed in the previous sections, being more energetically favourable.  

\section{Conclusions}
\label{conclusions}

In this paper {\color{black} we focused on the Hawking effect in dispersive dielectric media, and in particular 
our focus was on the analytical proof of thermally in the spectrum of emitted photons. Our reference 
framework was the Hopfield model, suitably modified in order to account phenomenologically of some 
characteristics of the Kerr effect. Several aspects of the model were also taken into account, and interesting results 
were obtained.} We sum up our results as follows.
\begin{enumerate}
\item The microscopical (or rather, mesoscopical) model apt to the phenomenological description of the process 
involved in thermal pair creation has been previously identified with the Hopfield model. In place of the model 
with a step-wise behavior of the refractive index \cite{finazzi-carusotto-kin,finazzi-carusotto-pra13,finazzi-carusotto-pra14}, 
we have developed a model where microscopic parameters are left free to vary smoothly in space-time coordinates.  
{\color{black} This model allows extensions to multi-resonances situations for the polarization field. 
We have also introduced a simpler model involving  
a scalar field doublet in place of the full electromagnetic field and of the polarization field, 
in order to get a simpler and more manageable model, which still preserves the same dispersion relation as the 
original one}. A conserved scalar product has been identified, in 
order to provide a norm for identifying particle and anti-particles states. {\color{black} In Appendix \ref{manley}, from Wronskian relations, 
the generalized Manley-Rowe relations are deduced in a scattering framework (they were deduced 
by other means in \cite{kravtsov,ostrovskii,sorokin})}. 
\item The analytical calculations {\color{black} for thermality have been based on matching the asymptotic 
solutions obtained in the WKB approximation and the solutions of the differential equation in Fourier space where an 
expansion up to the linear order of the refractive index has been performed. In particular, we have shown that 
calculations can be developed in a nice parallel way with respect to 
known calculations concerning fluid models \cite{corley,cpf}. We have also tried to identify a common denominator 
between the two aforementioned classes of models (i.e. fluid models and dielectric models). 
On one hand, if one 
pursues a second order expansion in Fourier space near the horizon, a Fuchsian singularity structure of the 
second order equation one obtains can be associated with Hawking effect. On the other hand, a first order 
expansion has been shown to be also very interesting (and sufficient for the aim to find out thermality),  
and in this case a Fuchsian singularity structure can be still identified, and a} phase factor with a logarithmic branch point has been identified as the source of thermality. 
In all cases, thermality is involved with a ternary process (a fourth state is shown to decouple from the spectrum, 
as in \cite{corley,cpf}). 
\item The aforementioned picture can be extended to the multi-resonances situation. Thermality is still 
preserved, and the possibility to get it in processes involving the lower branch and also the upper one 
has been indicated. We deserve to come back to this topic in future studies. 
\item A more tricky, and still open, problem concerns what should be meant by `horizon' in the near horizon 
expansion. We have discussed this problem in our model, and the non-dispersive horizon, the group horizon and also 
the phase horizon could have chances to be the right places fou our model. GH is a strong candidate. Our analysis cannot yet be conclusive. 
In particular, the expected role of a blocking horizon is not yet emerging in a neat way from the present (as well as in the existing one, albeit on a analytical footing) analysis. {\color{black} Furthermore, a detailed discussion, in comparison with the analysis of the recent paper 
\cite{coutant-broad}, where it is shown that in fluid models the wave function associated with the modes involved in the scattering process is not able to distinguish between different kinds 
of horizons, would be required in order to delve into this interesting problem.}
\end{enumerate}
Our analysis is completed in the appendix. {\color{black} In Appendix \ref{exact} we pointed out that the model, in line of principle, 
can be solved exactly in the general case, without necessarily referring to the physical situations involved in the 
Hawking effect. In Appendix \ref{scattering-asympt}, the asymptotic behavior of the solution was considered. 
In Appendix \ref{manley} the generalized Manley-Rowe identities were discussed, and in Appendix \ref{geomopt} some 
relevant aspects of the geometrical optics approximation were discussed too.} \\
Further theoretical studies and further experimental analysis are still required in order to give better and stronger grounds 
to this fascinating field of investigation. 

\acknowledgments
We thank Daniele Faccio for his suggestions and remarks, and for several discussions about the physics at hand.

\appendix

\section{The general theory: exact treatment}
\label{exact}
In the $\varphi\psi$-model,
we can also obtain in line of principle the exact dispersion relation. 
For an example, we deal with the case where $\chi=\chi (x-v t)$ in the lab frame (see also \cite{belcacciadalla-pert}).
In order to delve into the physical content of the above equations, let us introduce $\tilde{G}_\psi$, i.e. 
the Fourier transform of the Green function $G_\psi$. Then,  for $k=(\frac{\omega}{c},\vec{k})$, from equation (\ref{eq-varphi}) we obtain 
\beq
k^\mu k_\mu -\frac{g^2}{c^2} (v^\mu k_\mu)^2 \tilde{G}_\psi (k_\nu)=0,
\label{exact-disp}
\eeq
which represent the exact dispersion relation for the theory at hand. This relation holds for a generic 
space-time dependence of the susceptibility $\chi$, and even of the proper frequency $\omega_0$. It 
contains all the necessary information in order to explore the problem of pair-creation, as we shall see in the following.\\

 Let us 
consider $\chi=\chi (x)$ in the comoving frame: we obtain in 2D, with $v^\mu = \gamma (c,v)$ and $k^\mu=\gamma(\frac{\omega}{c},k)$,
\beq
\frac{\omega^2}{c^2}-k^2 -\frac{g^2}{c^2} \gamma^2 (\omega+v k)^2 \tilde{G}_\psi (\omega,k)=0.
\label{exact-comoving}
\eeq
This equation in the lab becomes 
\beq
k_{lab}^2 c^2 =\omega_{lab}^2 [1-g^2   \tilde{G}^{lab}_\psi (\omega_{lab},k_{lab})],
\label{exact-lab}
\eeq
where we define
\beq
\tilde{G}^{lab}_\psi (\omega_{lab},k_{lab})\equiv \tilde{G}_\psi (\gamma (\omega_{lab}-v k_{lab}),
\gamma (k_{lab}-\frac{v}{c^2} \omega_{lab})).
\eeq
Then, we can define the refractive index as
\beq
n^2  (\omega_{lab},k_{lab}) = 1 - g^2 \tilde{G}^{lab}_\psi (\omega_{lab},k_{lab}),
\label{ref-index}
\eeq
which is an exact relation for the refractive index in the travelling perturbation case with $v=$const. 
It is interesting to note that, when  no perturbation is present, one obtains 
\beq
n^2  (\omega_{lab},k_{lab}) = 1 + g^2  \chi_0 \omega_0^2 \frac{1}{\omega_0^2 -\omega^2_{lab}},
\eeq
which has exactly the same form as for the standard Hopfield model (apart from a suitable re-definition of $g$). 
Indeed, we obtain easily 
\beq
\tilde{G}^{lab}_\psi (\omega_{lab},k_{lab}) =  \frac{-\chi_0 \omega_0^2 }{\omega_0^2 -\omega^2_{lab}}
\eeq
in the aforementioned case. It is to be noted that (\ref{exact-comoving}) depends on the explicit form 
of $\chi (x)$ in the comoving frame, and that in general the refractive index in the lab frame is expected to 
display a dependence also on $k$, i.e. also spatial dispersion is to be expected in the general case. 

In the 2D  case, one can also write formally the exact solutions for the equation of motion as follows:
\beqnl
\varphi (t,x) &=& \int \frac{dk d\omega}{(2\pi)^2} \delta (\frac{\omega^2}{c^2}-k^2 - 
\frac{g^2}{c^2} \gamma^2 (\omega+v k)^2 \tilde{G}_\psi (\omega,k)) \tilde{\varphi} (\omega,k) e^{i\omega t + i k x},\\
\psi (t,x) &=& g^2 \int \frac{dk d\omega}{(2\pi)^2} \delta (\frac{\omega^2}{c^2}-k^2 - 
\frac{g^2}{c^2} \gamma^2 (\omega+v k)^2 \tilde{G}_\psi (\omega,k)) \tilde{G}_\psi (\omega,k) i \gamma (\omega+v k) 
\tilde{\varphi} (\omega,k) e^{i\omega t + i k x}.
\eeqnl

We are of course interested in quantum aspects. It is very useful to point out that the model 
is, at least in line of principle, exactly soluble. Indeed, if we define
\beq
\Phi:= \left(\begin{array}{c}
\varphi\\
\psi
\end{array}
\right),
\eeq
we can obtain the following form for the action:
\beq
S := \frac{1}{2} \int d^4 x \Phi^T {\mathcal Q} \Phi,
\eeq
where 
\beq
{\mathcal Q}:=\left[
\begin{array}{cc}
-\square & g  v^\alpha \partial_\alpha\cr
-v^\alpha \partial_\alpha g & -v^\alpha \partial_\alpha \frac{1}{\chi \omega_0^2} 
v^\beta \partial_\beta - \frac{1}{\chi}
\end{array} \right].
\eeq
As it stands, the action is quadratic in the field $\Phi$, and this implies that the quantum theory 
is exactly soluble, as the path-integral formalism immediately reveals. Any amplitude is related to the 
Green function of the matrix operator ${\mathcal Q}$, and, moreover, any (spontaneous) pair-creation 
process induced by the presence of a spacetime-dependent $\chi$ is in line of principle exactly 
calculable, being associated with the imaginary part of the effective action, which could be obtained, 
as usual, via $\zeta$-function techniques, after having calculated the effective action 
(which amounts to calculating the functional determinant of ${\mathcal Q}$). This route, even in simple 
cases can be very involved also for a Gaussian model as the one we are setting up. 
The approach we have adopted for revealing quantum instabilities, which consists in checking the presence of negative norm states 
(antiparticles) in stimulated scattering, is simpler. 
Sometimes, this approach is called transmission coefficient approach \cite{damour}.

\section{Asymptotic bases}
\label{scattering-asympt}

We consider in what follows the asymptotic behavior (in x) for simplicity in the case of the 2D model 
(in the 4D case, separation of variables also on transverse variables allows to show that only  
a little and not substantial modification of the calculations displayed below occurs; see also \cite{belcacciadalla-hopfield}). 
We work on stationary solutions, so we get the following second order system of ordinary 
differential equations:
\beqnl
\psi'' &=& ((\log \chi \omega_0^2)' + 2 i \frac{\omega}{v}) \psi' +
\left[-i \frac{\omega}{v} (\log \chi \omega_0^2)'-\frac{\omega_0^2}{\gamma^2 v^2}) +
\frac{\omega^2}{v^2}\right] \psi -i \frac{g}{c} \chi \omega_0^2 \frac{\omega}{\gamma v^2} \varphi-
\frac{1}{\gamma v c} \chi \omega_0^2 \partial_x (g \varphi),\\
\varphi'' &=& \frac{g}{c} \gamma v \psi' -i \frac{g}{c} \gamma \omega \psi - \frac{\omega^2}{c^2}\varphi.
\eeqnl
Then, we associate with it a first order system by introducing 
\beq
p:=\psi', \quad \quad q:=\varphi'.
\eeq
Then, if $W(x):=(\psi (x), \varphi (x), p(x), q(x))^T$, we obtain the following first order system:
\beq
W'(x)=K_4 W(x),
\eeq
where the $4\times 4$ matrix operator $K_4 (x)$ has the following structure:
\beq
K_4:=\left[ \begin{array}{cc}
0_2 &  1_2 \cr
A_2 & B_2 
\end{array}
\right].
\eeq
$0_2,1_2$ are $2\times2$ matrices, the first one with all entries equal to zero, 
and the second one is the identity. As to $A_2,B_2$, we have
\beq
A_2:=\left[ \begin{array}{cc}
\frac{\omega^2}{v^2} -i \frac{\omega}{v} (\log \chi \omega_0^2)'-\frac{\omega_0^2}{\gamma^2 v^2} 
& 
-i \frac{g}{c} \frac{\omega}{v} \frac{\chi \omega_0^2}{\gamma v}  -\frac{\chi \omega_0^2}{\gamma v c} g'\cr
-i \frac{g}{c} \gamma \omega 
& 
- \frac{\omega^2}{c^2}
\end{array}
\right],
\eeq
and
\beq
B_2:=\left[ \begin{array}{cc}
2 i\frac{\omega}{v} +(\log \chi\omega_0^2)' 
& 
- \frac{g}{c}  \frac{\chi \omega_0^2}{\gamma v}  \cr
\frac{g}{c} \gamma v 
& 
0
\end{array}
\right]. 
\eeq
Let us write
\beq
K_4 = {\mathcal C}+{\mathcal R},
\eeq
where ${\mathcal C}$ is a constant matrix and ${\mathcal R}={\mathcal R}(x)$: 
\beq
{\mathcal C}:=\left[ \begin{array}{cc}
0_2
& 
1_2\cr
A_c 
& 
B_c
\end{array}
\right], 
\eeq
with 
\beq
A_c:=\left[ \begin{array}{cc}
\frac{\omega^2}{v^2}-\frac{\omega_0^2}{\gamma^2 v^2} 
& 
0\cr
0 
& 
- \frac{\omega^2}{c^2}
\end{array}
\right],
\eeq
and
\beq
B_c:=\left[ \begin{array}{cc}
2 i\frac{\omega}{v} 
& 
0 \cr 
0 
& 
0
\end{array}
\right];  
\eeq
moreover, 
\beq
{\mathcal R}:=\left[ \begin{array}{cc}
0_2
& 
0_2\cr
A_r 
& 
B_r
\end{array}
\right], 
\eeq
with 
\beq
A_r:=\left[ \begin{array}{cc}
-i \frac{\omega}{v} (\log \chi \omega_0^2)' 
& 
-i \frac{g}{c} \frac{\omega}{v} \frac{\chi \omega_0^2}{\gamma v} - \frac{\chi \omega_0^2}{\gamma v c} g'\cr
-i \frac{g}{c} \gamma \omega
& 
0
\end{array}
\right],
\eeq
and 
\beq
B_r:=\left[ \begin{array}{cc}
 (\log \chi \omega_0^2)'
& 
- \frac{g}{c}  \frac{\chi \omega_0^2}{\gamma v}  \cr
\frac{g}{c} \gamma v
& 
0
\end{array}
\right].  
\eeq
Under the hypothesis 
\beq
\int_a^\infty dx |{\mathcal R}(x)|<\infty,
\eeq
which physically can match very well the nature of travelling perturbation of $\delta n$, to be implemented by means of 
a suitable choice of the microscopic parameters $g,\omega_0,\chi$,  
according to theorems in \cite{eastham}, we can infer that, both as $x\to \infty$ and as $x\to -\infty$, 
the asymptotic behavior of solutions is governed by the eigenvalues of ${\mathcal C}$, which implies that 
the basis for the homogeneous case with $g,\omega_0,\chi$ asymptotically constants, is
 asymptotically a good scattering basis also for the perturbed problem.  
To be more precise: the asymptotic region solutions are a scattering basis, and, moreover, solutions of the 
full equations asymptotically behave as the asymptotic region solutions, which then represent a good 
scattering basis. Furthermore, we are interested in (localized) wave-packets, whose support is finite.  
This is relevant as far as we are concerned with the problem of defining particle and antiparticle states.

\section{Generalized Manley-Rowe relations and pair creation amplitudes}
\label{manley}

We give a more systematic account of pair-creation amplitudes, which can be used both for analytical calculations 
and for numerical ones (see in the latter case results in \cite{rubino-njp,rubino-sr,petev-prl,rubino-phd-thesis}). At first, we focus on the two-dimensional problem and 
we fix a scattering basis in the asymptotic regions. At fixed $\omega$ we have a number of states as $x\to -\infty$, 
which is equal to the number of intersections between the horizontal line $\omega=$const and the asymptotic dispersion 
relation $\omega=f(k)$, see Figure \ref{fig:asymptDR}. Some of them can have positive group velocity $v_g>0$ (right moving), and some others can have 
$v_g<0$ (left moving). In situations where a blocking horizon is present, only the states  
 belonging to the asymptotic region on the left (with $\delta n=0$) are involved. In situations where 
blocking is absent or frequencies involved do not admit blocking, then also (transmitted) scattering states 
belonging to the asymptotic region on the right (with $\delta n \not=0$) are involved. See also \cite{finazzi-carusotto-pra13,
finazzi-carusotto-pra14}.\\

Some preliminary considerations are in order, concerning the Hawking effect. We shall discuss 
the scattering process involved in the Hawking effect for a white hole in this section, whereas in Section  \ref{thermality} thermality is discussed for a black hole. There is no contradiction, as white hole is the 
time-reversal of a black hole. Still, at the level of scattering, one has to investigate what happens. 
Time-reversal implies $(\omega,v)\mapsto (-\omega,-v)$. 
We notice that the original Hopfield model, and also the one with varying $\chi,\omega_0,g$, is invariant under time reversal, as the Lagrangian of the model is. As to the $\varphi \psi$-model we have introduced, 
equations of motion are invariant under time-reversal provided that $g\mapsto -g$. The latter freedom 
for the scalar model can be assumed without any problem, provided that the correct branch for the relation 
between microscopic parameters and macroscopic ones is chosen. \\

In general, we can expect to deal with several branches of the dispersion relation. In that case, it may be more 
useful to consider the asymptotic dispersion relations in the lab. An analogous reasonings leads to 
a number of states as $x\to \infty$, for which again $v_g>0$ or $v_g<0$. A complete scattering basis is obtained 
by considering both a scattering with one initial right moving state, which can give rise to several 
reflected states and one transmitted state, and a scattering with one initial left moving state, with an 
analogous behavior. This is particularly important for the actual computation of the pair creation amplitude in 
the spontaneous emission case.\\
Amplitudes can be calculated both in the traditional framework, by means of Bogoliubov transformations, 
or by means of the conservation of fluxes in the scattering process in the comoving frame. We adopt the latter frame, 
which is also more directly related to some previous works in literature \cite{kravtsov,ostrovskii,sorokin}. 
We know that, in the comoving frame, there is a current $J_x$ which is conserved, as we have shown in
Section \ref{hopf-model}.
In particular, we can consider $J_x$ as a bilinear form:
\beq
J_x (\Psi_1,\Psi_2),
\eeq
where $\Psi_1, \Psi_2$ is a couple of asymptotic plane wave solutions of the equations of motion. 
In the scattering `$\rightarrow$', with a single initial state which is right moving, scattering solutions 
are denoted by $\Psi_{\rightarrow}$. An analogous definition is given for the scattering `$\leftarrow$' and 
$\Psi_{\leftarrow}$. Then we can obtain a number of `wronskian relations', for example we 
can calculate 
\beq
J_x (\Psi_{\rightarrow}^\ast,\Psi_{\rightarrow}).
\label{wro1}
\eeq
For definiteness, let us consider, for the case $IN\to P+N^\ast+B+T$, where $T$ stays for a possible 
transmitted state, the following state:
 \beq
\Psi_{\rightarrow} = N_\Psi e^{-i \omega t} 
\begin{cases}
T_{IN}^{\rightarrow} W_{IN} e^{i k_{IN} x} + 
T_{P}^{\rightarrow} W_{P} e^{-i k_{P} x} +
T_{N^\ast}^{\rightarrow} W_{N^\ast} e^{-i k_{N^\ast} x} +
T_{B}^{\rightarrow} W_{B} e^{-i k_{B} x}
  &\quad \hbox{for}\ x\to -\infty,\cr
T_{T}^{\rightarrow} W_{T} e^{i k_{T} x} 
&\quad \hbox{for}\ x\to \infty,
\end{cases}
\eeq
where $T_{IN}^{\rightarrow},T_{P}^{\rightarrow},T_{N^\ast}^{\rightarrow},T_{T}^{\rightarrow},T_{B}^{\rightarrow}$ are the usual 
scattering coefficients with the additional label $\rightarrow$ indicating that the initial state is right 
moving, and where  $ W_{IN}$ etc. are vector Fourier components of the plane wave which is considered. 
Then, (\ref{wro1}) is of the following form: 
\beq
1 -|P|^2 -|B|^2 -|T|^2 + |N|^2 =0.
\label{mr-gen}
\eeq
Interference terms are washed out asymptotically in time, in the sense that separated `photon packets' for the 
various modes are expected on long time scales (long with respect to the interaction time scale).  
This emerges from current conservation:
\beq
J_x^{left}=J_x^{right},
\eeq
where `left' and `right' indicate states on the left and on the right of the step-like 
potential as $x\to -\infty$ and $x\to \infty$ respectively. For well-separated packets, we also 
get
\beq
J_x^{IN}+J_x^{P}+J_x^{N^\ast}+J_x^{B}=J_x^{T}.
\eeq
The above quantities have a sign which is determined by
\beq
J_x^U = \sign (v_g^U) |J_x^U|,
\eeq
where $U$ is meant as a positive norm asymptotic solution; furthermore, we take into account that 
\beq
J_x^{N^\ast}=-J_x^N,
\eeq
i.e. the antiparticle state (negative norm) current is opposite to the corresponding particle state 
(positive norm) current. Cf. also \cite{damour}. 
Then, we define the following quantities:
\beq
\frac{J_x^U}{J_x^{IN}}=: \sign (\frac{v_g^U}{v_g^{IN}}) |U|^2, \quad \quad U=P,B,T,N.
\eeq
It is not difficult to show that 
\beq
|U|^2 = F(\omega,k_U) |T_U^{\rightarrow}|^2, \quad \quad U=P,B,T,N
\eeq
where $F (\omega,k_U)$ is a positive kinematic coefficient which of course depends on the current structure. 
This analysis can be easily extended to the case of an arbitrary number of states (compatibly with the dispersion relation).\\
It is also remarkable that, in the spontaneous case, we have to consider
\beq
<J_x>:=<0|J_x|0>=\sum_{\rightarrow} \sum_{\leftarrow} \int 
 \frac{d\omega}{(2\pi)} \frac{dk}{(2\pi)} \frac{1}{N_\Psi} \delta (\textrm{dispersion relation}) V^\ast V,
\eeq
where the sum has to be extended both to initial left-moving states and to initial right moving ones. This 
leads to the same particle creation amplitudes associated with (\ref{mr-gen}).\\
As to the conservation of the fluxes, we recall that we could also define the Poynting vector for the theory at hand. 
It is easy to conclude that the same amplitudes as above would be obtained.

\section{Geometrical Optics}
\label{geomopt}

The eikonal approximation is a usual tool for analyzing solutions in the framework of 
analogue gravity. We explore also this conceptual frame, because it can give useful 
suggestions and provide us also analytical tools for a better comprehension of the 
phenomenon at hand. In particular, it is remarkable that thermality of the Hawking radiation 
arises as associated with the presence, in the comoving frame, of a so-called group horizon 
(GH), i.e. a turning point (TP) for the waves which reach the perturbation, at least for frequencies in a given interval. E.g. in WKB approximation, a turning point is to be handled with care, due to the fact that it violates the requirements of the approximation itself. In geometrical optics, a TP represents a caustic for rays, so, again, the eikonal approximation fails there. It is necessary to point out immediately the limits of the given approximation, due to the fact that all the phenomenology 
which we are interested in arises near such a TP, where some other analytical tool has to be assumed.\\
In what follows, we  point out that even in presence of dispersion, the 
eikonal approximation still gives useful suggestions, and the method of characteristics can be 
used in order to explore solutions (geometrical optics is a good tool for studying 
the problem in the non-dispersive case, as known). Moreover, the problem of the group horizon, and also the 
problem of the phase horizon, can be exactly solved in the Cauchy approximation. \\

We shall limit ourselves mainly to the 2D eikonal equation 
\beq
\omega_{lab} n(\omega_{lab}, x-vt) = \pm c k_{lab}.
\eeq 
Let us consider the Cauchy approximation (\ref{cau-n}). We have in the comoving frame \cite{belgiorno-prd}:
\beq
G=0 \Longleftrightarrow (\omega + v k) (n(x)+B \gamma^2 (\omega+ v k)^2) - c k -\frac{v}{c} \omega=0.
\label{drel-cau}
\eeq
We are interested in the expression for the 
group horizon (if any), which is obtained by solving the system \cite{belgio-tunopt}
\beqnl
G&=&0,\\
\pa_k G&=&0.
\eeqnl
As to the latter equation, we obtain
\beq
\pa_k G=0 \Longleftrightarrow 3 B \gamma^2 v (\omega + v k)^2-v \left( \frac{c}{v}-n(x) \right)=0,
\label{gh-cau-eq}
\eeq
which can be solved explicitly:
\beq
(\omega + v k) = \pm \left( \frac{\frac{c}{v}-n(x)}{3 B \gamma^2}\right)^{1/2}.
\eeq
By substitution of the positive root in $G=0$, as we mean to get the group horizon for positive norm 
waves, we obtain an equation for $n(x)$ which 
allows us to find out explicitly the group horizon:
\beq
\frac{c}{v}-n(x) = 3 B \gamma^2 \left( \frac{1}{2 B \gamma^4} \frac{c}{v} \right)^{2/3} \omega^{2/3}=: \zeta_B \omega^{2/3},
\label{gh-cau-explicit}
\eeq
where $\zeta_B \propto B^{1/3}$. So we are able to find out $x_{GH}(\omega)$, which is a function of the frequency $\omega$, 
as expected.\\
As to the phase horizon, we have in the comoving frame that it corresponds to $\omega=0$. By taking into account the 
dispersion relation (\ref{drel-cau}), we find that 
\beq
\frac{c}{v}-n(x) = B \gamma^2 v^2 k_0^2,
\label{ph-eq}
\eeq
where $k_0$ is the value at which the dispersion relation $G=0$ intersect the $k$-axis (i.e. $\omega$=0). 
In the latter case, $x_{PH}(k_0)$ is a function of the aforementioned parameter, which is independent 
from $\omega$.

\end{document}